\documentclass[aps,prl,preprint,superscriptaddress,showpacs,floatfix]{revtex4}

\usepackage{graphicx}
\usepackage{bm}

\begin{document}


 \title{Using Available Volume to Predict Fluid Diffusivity \\ in Random
 Media}


 \author{Jeetain Mittal}
 \email[]{jeetain@che.utexas.edu}
 \affiliation{Department of Chemical Engineering, The University of Texas at 
 Austin, Austin, TX}

 \author{Jeffrey R. Errington}
 \email[]{jerring@buffalo.edu}
 \affiliation{Department of Chemical and Biological Engineering, University at 
 Buffalo, The State University of New York, Buffalo, NY}

 \author{Thomas M. Truskett$^{1,}$}
 \email[]{truskett@che.utexas.edu}
 \affiliation{Institute for Theoretical Chemistry, The University of Texas at Austin, Austin, TX}


 \date{\today}

\begin{abstract}
We propose a simple equation for predicting self-diffusivity of fluids
embedded in
random matrices of identical, but dynamically frozen, particles (i.e.,
quenched-annealed systems).  The only nontrivial input is the
volume available to mobile particles, which also can be
predicted for two common matrix types that 
reflect equilibrium and
non-equilibrium fluid structures.  The proposed equation can account
for the large differences in mobility exhibited by 
quenched-annealed systems with
indistinguishable static pair correlations, 
illustrating the key role that available volume plays 
in transport.
\end{abstract}

\maketitle
%
Fluid transport in random porous media is central to
a host of natural phenomena and technological applications, from
the function of biological cells to the performance of materials for 
membrane separations and heterogeneous catalysis.  
Although it is now appreciated that kinetic processes
in these systems are intimately connected to their 
microstructures (pore volume, surface area, 
connectivity, etc.)~\cite{torquato2002}, determining the precise
structure-property relationship for the transport property of
interest remains a formidable challenge.

One promising line of inquiry has been the exploration of
simple model systems for which both transport properties and
structure are amenable to theoretical analysis. 
In the case of diffusivity, earlier work has focused primarily on 
single particle
transport through random configurations of 
static obstacles~\cite{Bunde1991}.
At low obstacle densities, particles show anomalous, sub-diffusive
motion over intermediate time and length scales but recover normal diffusive
behavior in the long time limit.  On the other hand, sub-diffusive 
motion is observed for 
all times at high obstacle densities.  Similar
behavior is predicted for transport of single ions in disordered 
matrices of quenched charges~\cite{Deem1994}.    

Self-diffusivities of highly coupled fluids in random media pose additional
challenges and are also of great interest.  Very recently, 
insightful molecular dynamics simulation~\cite{Chang2004} 
and mode-coupling theory~\cite{qa-mct} studies have been presented for 
the hard-sphere (HS) fluid embedded in disordered matrices of 
obstacle spheres.  This type of model, introduced
originally by Madden and Glandt~\cite{MADDEN1988},
is referred to as a quenched-annealed (QA) system.  
To the best of our knowledge, no simple theoretical relationship between
self-diffusivity of mobile particles, densities of mobile and 
matrix particles, and
matrix microstructure has been proposed for QA media.  As a result,
a general consensus on which physical parameters are most 
important for understanding mobility in these model porous materials
is lacking. 

In this Rapid Communication, 
we introduce one such approximate relationship that
we motivate by using a physical argument for how the matrix 
reduces the volume available for 
diffusion of the mobile particles.
We test its predictions against the 
numerical results of molecular dynamics simulations for
HS QA systems with matrices that reflect equilibrium and
non-equilibrium structures, which serve as idealizations for two
different classes of physically realizable materials.  We then demonstrate 
how the available volume in both types of matrices can be 
accurately predicted directly from equilibrium properties of the 
bulk HS fluid.
Finally, we test our equation for diffusivity against simulation
results of 
QA systems comprising Lennard-Jones (LJ) particles.  
For all models investigated, we find good agreement with 
simulations over a wide range of parameters, 
with significant overpredictions of diffusivity occuring 
only for high matrix densities where limited connectivity 
of the available volume (not accounted for in our approximate
equation) also hinders transport.

The protocols that we use to generate static matrices represent 
two of the most commonly employed in studies of QA 
systems~\cite{Chang2004,MADDEN1988,Van1997a}.
In the first, $N_{\text M}$ particles of diameter $\sigma$ 
are initially equilibrated in a volume $V$ at temperature $T$.  
They are subsequently
quenched (i.e., frozen in place) in an equilibrium configuration, 
and a ``fluid'' (F) of $N_{\text F}$ identical 
particles is added, equilibrated, and 
studied.  This method of matrix generation is referred to as QA-M because
the matrix (M) itself 
reflects an equilibrium configuration of density 
$\rho_{\text M} = N_{\text M} \sigma^3/V$ for the bulk system.
QA-M matrices serve as elementary models of 
amorphous solid materials prepared, e.g, by very rapidly cooling
gel-forming suspensions of proteins or colloids~\cite{Van1997a}.

The second matrix generation protocol we study 
is referred to as QA-FM because it involves first equilibrating
$N_{\text F} + N_{\text M}$ identical particles of diameter $\sigma$
in a volume~$V$ 
at temperature~$T$.  Then, $N_{\text M}$ of the particles are
randomly selected and quenched to create the solid matrix.
The other $N_{\text F}$ particles constitute the mobile 
``fluid'' of density $\rho_{\text F} = N_{\text F} \sigma^3 /V$. 
QA-FM matrices are idealized
models for templated porous solids synthesized by depleting a
high density material of 
one of its components by, e.g., dissolution, reaction, 
or desorption~\cite{Van1997a}.  

Our approach for predicting the fluid self-diffusivities of 
these systems is motivated by three basic observations about their
behaviors.  
(i)~The total pair correlation functions of HS QA-M and QA-FM
configurations, averaged over mobile and matrix particles together at
the same total density $\rho_{\text F}+\rho_{\text M}$, are 
indistinguishable~\cite{Chang2004}.
(ii)~Yet,
their fluid self-diffusivities generally 
differ significantly, even when compared at the same ratio of 
matrix to mobile particles.  This has previously 
been interpreted as evidence that the differences in the
dynamics of QA media
cannot be predicted based on static structural information 
alone~\cite{Chang2004}.  (iii) However, there is a key 
static property that distinguishes individual QA systems:  the fraction
of volume available to the mobile particle centers 
in the matrix, $V_0/V$~\cite{Van1997a}.  As we demonstrate below, 
differences in this quantity can largely account for 
the wide range of self-diffusivities exhibited by QA materials
that are otherwise ``structurally similar''.

For example, contrast the behavior of a bulk HS fluid with
density $\rho=\rho_{\text F}+\rho_{\text M}$ to a HS
QA system (produced by either -M or -FM protocols) with matrix and 
mobile particle densities of $\rho_{\text M}$ and $\rho_{\text F}$, 
respectively.  The binary collisions that mediate diffusion 
in these two systems are expected to be comparable, since their 
pair correlation functions and mobile-particle thermal velocities 
are the same.  
Nonetheless, mobile particles in
the QA system will diffuse over a shorter characteristic
length scale per unit time compared to those of the bulk because the
volume available for diffusion in the QA matrix,
$V_0$, is smaller than $V$.  If we assume that the characteristic
length scales for diffusion are proportional to the cube root of 
the respective
available volumes and, in turn, that
the associated 
self-diffusivities are proportional to the square of the length
scales, 
we arrive
at the following approximate relationship:
\begin{equation}
\label{simpleD}
D(\rho_{\text F}, \rho_{\text M}) \approx D(\rho_{\text F}+
\rho_{\text M},0) \times (V_0/V)^{2/3},
\end{equation}
where $D(\rho_{\text F}, \rho_{\text M})$ and $D(\rho_{\text F}+
\rho_{\text M},0)$ are the self-diffusivities of the mobile particles 
in QA and bulk fluid systems. 
As we show below, the appeal of this relation is its ability
to predict the self-diffusivity of non-trivial QA systems from knowledge of 
well-characterized equilibrium properties of the {\em bulk} HS fluid.

In HS QA-M systems, matrices are drawn from
equilibrium configurations of the 
bulk HS fluid with density $\rho_{\text M}$, and thus 
$V_0/V$ is given by~\cite{reiss}
\begin{equation}
\label{reiss}
V_0/V = \exp [-\mu^{\text {ex}} (\rho_{\text M})/k_{\text B}T].
\end{equation}  
The excess chemical potential (relative to ideal gas) 
of the HS fluid, $\mu^{\text {ex}}$, 
can be obtained analytically from, e.g., either
scaled particle theory~\cite{reiss} or the Carnahan-Starling (CS)
equation of state~\cite{Carnahan1969} (here, we adopt the latter,
which is very accurate for $\rho_{\text M} < 0.9$).

In HS QA-FM media, however, the matrix structure is different than 
that of an
equilibrium HS fluid because it is quenched in the presence of 
mobile particles.
Nonetheless, as Van Tassel and co-workers~\cite{Van1997a,Van1997} 
have demonstrated, because
of the specific protocol by which ``non-equilibrium'' QA-FM matrices are
created, one can still apply equilibrium liquid-state approaches 
(e.g., integral equation theories) 
to estimate $V_0/V$.  Here, we introduce an accurate information-theory
(IT) based stategy for accomplishing this task.  
As we show, its main advantage is that it
only requires knowledge of $\rho_{\text M}$, $\rho_{\text F}$, 
and the pair correlation function $g(r)$ of the bulk HS fluid.

\begin{figure}[h]
{\includegraphics{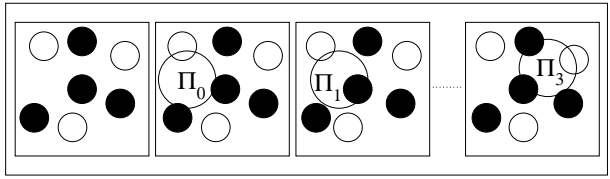}}
\caption{\label{Schematic} Schematic of a QA system. Equi-sized 
filled and empty circles
  are matrix and fluid particles, respectively. The bigger circle is a
  window with radius equal to 
the one particle diameter. $\Pi_i$ is the probability
that it will be occupied by exactly $i$ matrix or fluid particle centers.}
\end{figure}

Our strategy implicitly uses the fact that 
the pair correlations of the HS QA-FM system 
are indistinguishable from those of an equilibrium HS 
fluid with density $\rho_{\text F}+\rho_{\text M}$.  Since the
QA-FM particle identities (matrix or mobile) are randomly distributed,
one can write down an exact expression for $V_0/V$ in the QA-FM matrix 
in terms of equilibrium properties of bulk HS fluid:
\begin{equation} 
\label{Prob}
V_0/V = \sum_{i=0}^{\infty} \Pi_i(\rho_{\text M}+
\rho_{\text F}) \times
\left[1+\frac{\rho_{\text M}}{\rho_{\text F}}\right]^{-i}.
\end{equation}
Here $\Pi_i(\rho_{\text M}+\rho_{\text F})$ represents the
probability that a randomly placed spherical window (with radius
equal to one particle diameter) in a bulk HS fluid
of density $\rho_{\text F}+ \rho_{\text M}$ will contain precisely $i$ 
particle centers (see Fig.~\ref{Schematic}).  The quantitiy $[1+\rho_{\text M}/\rho_{\text F}]^{-i}$ 
is the probability that, in the structurally equivalent QA-FM system, all of the $i$ centers in the window would be 
fluid particles. 

\begin{figure}[h]
{\includegraphics{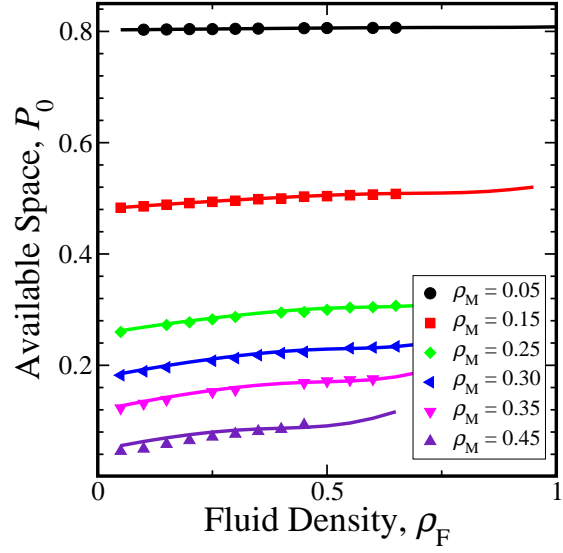}}
\caption{\label{IT-FM}The fractional available volume $V_0/V$ 
in HS QA-FM matrices
computed using the IT approach of Eq.~\ref{Prob} and
\ref{ProbN} (curves)
and ``exact'' results (filled circles) 
using the available space algorithm of Ref~\cite{Sastry1997a}.}
\end{figure}

IT provides expressions for the $\Pi_i$, 
\begin{eqnarray}
\label{ProbN}
\Pi_i=\exp [\lambda_0 + \lambda_1 i + \lambda_2 i^2]/i!,
\end{eqnarray}
which maximize a relative information entropy subject to some experimental 
constraints~\cite{Hummer1998}.
In particluar, $\lambda_0, \lambda_1,$ and $\lambda_2$ are 
Lagrange multipliers determined~(see, e.g., \cite{Hummer1998}) 
by imposing the normalization condition 
$\sum_i \Pi_i = 1$ and the first two moments of the window occupancy, 
$\overline{i}$ and $\overline{i^2}$,
\begin{eqnarray}
\label{Hill}
\overline{i} &=& 4 \pi \rho /3, \nonumber \\
\overline{i^2} &=& \overline{i} + \rho^2 \int_{\text w} d{\bf r} \int_{\text w} d{\bf r}'
g(|{\bf r}-{\bf r}'|). 
\end{eqnarray}
Here, $\rho = N \sigma^3/V$, and the subscript w 
in the last expression indicates that the
integrals are constrained to the spherical observation 
window.  Fig.~\ref{IT-FM} shows a numerical comparison of 
$V_0/V$ predicted by Eq.~\ref{Prob} and \ref{ProbN} 
to the ``exact'' results
for $V_0/V$ obtained from applying the available space 
algorithm of Sastry et al.~\cite{Sastry1997a} to HS QA-FM matrices. 
As is evident, the simple IT approach provides accurate
predictions over a wide range of matrix parameters.

Having established means for estimating $V_0/V$ of both 
QA-M and QA-FM matrices, we now only require an expression
for the self-diffusivity of the bulk HS fluid, $D(\rho,0)$, in order 
to predict HS QA diffusivity using Eq.~\ref{simpleD}.  For this, we
adopt Speedy's empirical fit to molecular dynamics simulation 
data~\cite{Speedy1987},
\begin{equation}
\label{speedyD}
D(\rho,0) = \frac{A}{\rho}\left(1 - \frac{\rho}{1.09}\right)
\left[1 + \rho^2(0.4 - 0.83\rho^2)\right],
\end{equation}
where $A=3/(8\sqrt{\pi})$.   We
implicitly non-dimensionalized $D(\rho,0)$ of Eq.~\ref{speedyD},
and all other self-diffusivities in this study,
by $\sigma^2/\tau$, where $\tau=\sigma\sqrt{m/k_{\text B}T}$, 
$k_{\text B}$ is the Boltzmann's constant, and $m$ is particle mass. 

To test the validity of Eq.~\ref{simpleD}, we performed 
molecular dynamics simulations~\cite{rap} for HS QA-M and QA-FM systems 
in the microcanonical
ensemble using $N_{\text F}=1500$ mobile particles and a
periodically-replicated simulation cell.  Runs were performed at
different values of $N_{\text M}$ and 
$V/\sigma^3$ to obtain results
for specific combinations of $\rho_{\text F}$ and $\rho_{\text M}$.
We extracted self-diffusivity of 
the mobile particles $D(\rho_{\text F}, \rho_{\text M})$ 
by fitting the long-time ($t \gg 1$) behavior of the 
average mean-squared 
displacement to the 
Einstein relation $\left<\Delta {\bf r}^2\right> = 6Dt$.

\begin{figure}[h]
{\includegraphics{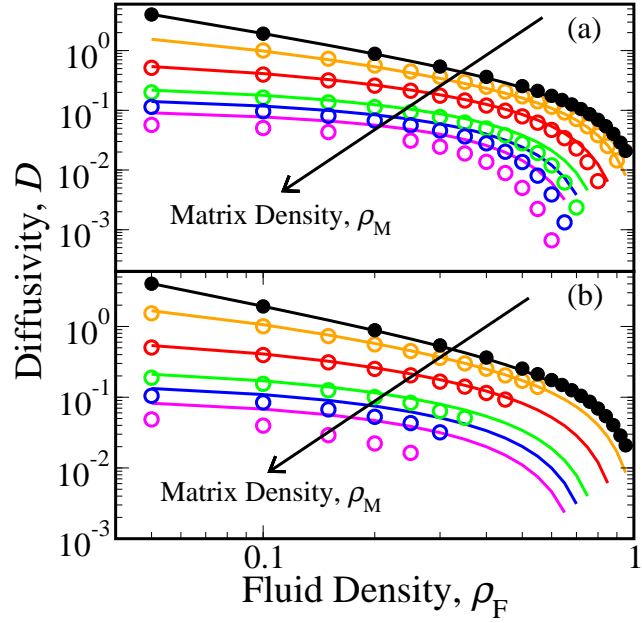}}
\caption{\label{D-HS} Fluid self-diffusivity $D$ versus fluid density 
$\rho_{\text F}$ for (a) HS QA-FM and (b) HS QA-M systems with matrix
densities $\rho_{\text M}=0.0$, 0.05, 0.15, 0.25, 0.30, and 0.35. Curves are 
predictions of Eq.~\ref{simpleD} and circles are molecular 
dynamics simulation results.} 
\end{figure}

Fig.~\ref{D-HS} provides a fairly comprehensive comparison of 
$D(\rho_{\text F},\rho_{\text M})$ from
Eq.~\ref{simpleD} to the 
results of our molecular dynamics simulations.  
Interestingly, the predictions show
semi-quantitative agreement with simulations 
of QA systems for matrix densities
in the wide range $0 \le \rho_{\text M} \le 0.25$, which includes systems
where matrix particles exclude mobile particle centers 
from over $70\%$ of the total volume (see, e.g., Fig.~\ref{IT-FM}).  Given
that matrix particles reduce $D(\rho_{\text F},\rho_{\text M})$ by
more than an order of magnitude across this range of densities, the success of 
Eq.~\ref{simpleD} argues that available volume plays
a primary role in controlling the
single-particle dynamics.  For matrix densities greater than 
$\rho_{\text M}=0.25$ (i.e., systems with less than $30\%$ of volume 
available to mobile particle centers), Eq.~\ref{simpleD} captures the qualitative
trends, but it systematically overpredicts the simulated 
self-diffusivities.  This overprediction is, of course, expected, given 
that Eq.~\ref{simpleD} only accounts for the reduction of available
volume and not the fact that available volume also becomes
highly disconnected (i.e., some pockets of available volume are
inaccessible or are accessible by only small number of paths) at high matrix densities~\cite{torquato2002}, which 
acts to further hinder transport.  However, for systems
of physical relevance (porous catalysts, membranes
for separations, transport gels for drug delivery, etc.) which rely on
rapid diffusion for their functionality, the precise connectivity
of the available volume will be less important.

In order to test whether the connection between diffusivity and 
available volume holds more generally for fluids confined to random 
media, we also
performed molecular dynamics simulations of QA-FM systems comprising 
LJ
particles (truncated and shifted with a quadratic function in
$r$~\cite{ford} to 
insure that both the potential and its gradient vanish at $r_{\mathrm
  {cut}} = 2.5$).  The details of the simulations are identical
to those of the HS QA systems described earlier, except that the
equations of motion were integrated via the velocity Verlet 
algorithm~\cite{rap},
and $N_{\text F} = 1000$ mobile particles were considered.
The goal was to test if one could 
employ techniques commonly used in thermodynamic 
perturbation theory to map the QA LJ system onto an 
equivalent QA HS system, and then use Eq.~\ref{simpleD} to predict
the self-diffusivity.  The specific mapping that we used in this
study is a Boltzmann factor criterion~\cite{hsu,amotz}, which determines
the temperature-dependent ``effective'' HS diameter $\sigma(T)$ (and
hence the corresponding reduced matrix and fluid denisties,
$\rho_{\text M}$ and $\rho_{\text F}$) of the LJ system 
through the following relationship $u_0 (r=\sigma) =k_{\text B}T$, where
$u_0$ is the repulsive part of the Weeks-Chandler-Andersen 
decomposition of the pair potential~\cite{wca}.  
The hypothesis is that LJ QA systems will exhibit similar reduced 
diffusivities $D(\rho_{\text F},\rho_{\text M})$ as HS QA systems when
compared at the same values of $\rho_{\text F}$ and $\rho_{\text M}$.

\begin{figure}[h]
{\includegraphics{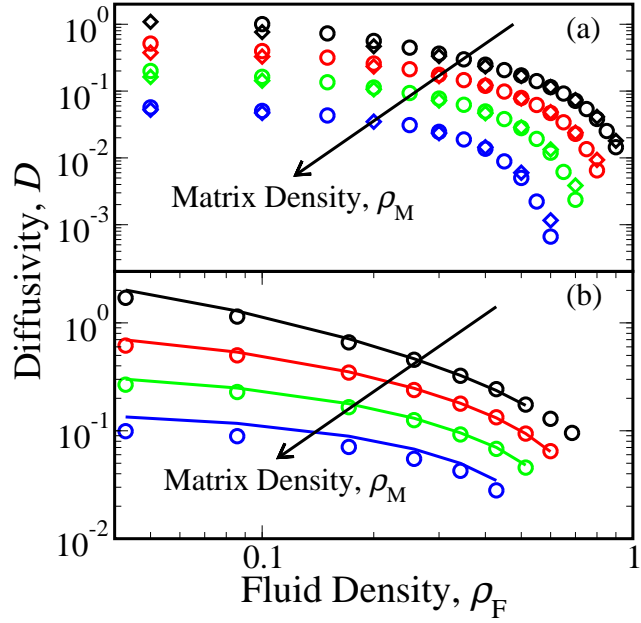}}
\caption{\label{D-LJ-HS}(a) Fluid self-diffusivity $D$ versus fluid density 
$\rho_{\text F}$ for HS QA-FM (circles) and LJ QA-FM (diamonds)
systems obtained via molecular dynamics simulations.  
The LJ QA-FM system is at 
$T=\epsilon/k_{\text B}$.  Matrix densities of $\rho_{\text M}=0.05$,
0.15, 0.25, and 0.35 are presented. (b) Results for the LJ QA-FM system at 
$T=3\epsilon/k_{\text B}$:  molecular dynamics simulations (circles) 
and Eq.~\ref{simpleD} (curves).  Matrix densities of $\rho_{\text M}=0.043$,
0.128, 0.214, and 0.3 are presented.
In both (a) and (b), matrix and mobile 
densities for the LJ QA-FM systems are defined as 
$\rho_{\text M}=N_{\text M} \sigma(T)^3/V$ and 
$\rho_{\text F}=N_{\text F} \sigma(T)^3/V$, respectively. 
The effective HS diameter $\sigma(T)$ is
determined by a Boltzmann factor criterion described in the text.}
\end{figure}

As a first test of this idea, we compare in Fig.~\ref{D-LJ-HS}a 
the diffusivities $D(\rho_{\text F},\rho_{\text M})$
of the HS QA-FM and LJ QA-FM systems obtained by our molecular
dynamics simulations.  In this plot, the LJ QA-FM system is at a temperature
$T= \epsilon/k_{\text B}$, where $\epsilon$ is the characteristic
energy of the LJ pair potential~\cite{ford}.  Except for
at very low fluid densities (conditions for which the LJ fluid structure is not accurately approximated by an equivalent HS reference fluid~\cite{edt}), 
the mapping brings the dynamics of the LJ and HS QA systems into
excellent agreement.

A final test is to check whether the simple aforementioned mapping 
allows direct prediction of LJ QA-FM dynamics at other temperatures
using Eq.~\ref{simpleD}.  Fig.~\ref{D-LJ-HS}b shows the comparison
of the predicted $D(\rho_{\text F},\rho_{\text M})$ versus the results
of our molecular dynamics simulations at $T= 3\epsilon/k_{\text B}$. 
The agreement between the predictions 
and the simulations again confirm the pivotal role that
available volume plays in controlling the single-particle dynamics of 
fluids in porous media. 

TMT and JRE acknowledge the financial support of the National
Science Foundation Grants No. CTS-0448721 and CTS-028772,
respectively, and the Donors of the American Chemical Society
Petroleum Research Fund Grants No. 41432-G5 and 43452-AC5, respectively.
TMT also acknowledges the support of the David and Lucile Packard and
Alfred P. Sloan Foundations.
The Texas Advanced Computing Center (TACC) provided computational 
resources for this study.

\newpage

\newpage
\newpage
\newpage

\end{document}